\documentclass[a4paper,review,11pt,authoryear]{elsarticle}
\usepackage{kpfonts}
\usepackage{amsmath,bm,hyperref,url,fullpage,mathtools,booktabs,parskip}
\usepackage{paralist,enumitem,todonotes,microtype}
\usepackage[a4paper,text={16cm,25cm}]{geometry}
\usepackage[labelsep=period]{caption}

\usepackage[section]{placeins}
\usepackage{algorithm}
\usepackage{algorithmicx}
\usepackage{algpseudocode}
\usepackage{verbatim}
\usepackage{multirow}
\usepackage{multicol}
\usepackage{arydshln}
\usepackage{color}
\usepackage{adjustbox}
\usepackage{amssymb}
\usepackage{tabularx}

\graphicspath{{IJFpaper-figure/}}

\usepackage{hyperref}
\hypersetup{
colorlinks=true,
linkcolor=blue,
citecolor=blue,
urlcolor=blue}

\providecommand{\tightlist}{%
  \setlength{\itemsep}{0pt}\setlength{\parskip}{0pt}}

\journal{International Journal of Forecasting}
\begin{document}
\date{}
\begin{frontmatter}

\title{Crude oil price forecasting incorporating news text}

\author[mainaddress]{Yun Bai \fnref{eqc}} \ead{baiyunbuaa@163.com}
\author[secondaryaddress]{Xixi Li \fnref{eqc}\corref{cor}}\ead{xixi.li@manchester.ac.uk}\ead[url]{https://orcid.org/0000-0001-5846-3460}
\author[mainaddress]{Hao Yu} \ead{yuhao1207@126.com}
\author[mainaddress]{Suling Jia} \ead{jiasuling@126.com}

\cortext[cor]{Corresponding author}
\address[mainaddress]{School of Economics and Management, Beihang
University, Beijing 100191, China.}
\address[secondaryaddress]{Department of Mathematics, The University of Manchester, Manchester M139PL, UK.}
\fntext[eqc]{The authors contributed equally.}

\begin{abstract}

Sparse and short news headlines can be arbitrary, noisy, and ambiguous, making it difficult for classic topic model \emph{LDA} (latent Dirichlet allocation) designed for accommodating long text to discover knowledge from them.
Nonetheless, some of the existing research about text-based crude oil forecasting employs \emph{LDA} to explore topics from news headlines, resulting in a mismatch between the short text and the topic model and further affecting the forecasting performance. Exploiting advanced and appropriate methods to construct high-quality features from news headlines becomes crucial in crude oil forecasting. To tackle this issue, this paper introduces two novel indicators of topic and sentiment for the short and sparse text data. Empirical experiments show that \emph{AdaBoost.RT} with our proposed text indicators, with a more comprehensive view and characterization of the short and sparse text data, outperforms the other benchmarks. Another significant merit is that our method also yields good forecasting performance when applied to other futures commodities.

\end{abstract}


\end{frontmatter}

\section{Introduction}
\label{introduction}
Crude oil is known as ``industrial blood'', due to the fact that the industry relies heavily on the supply of crude oil. Crude oil also plays an important role in the global economic system. Therefore, the accurate forecasting of crude oil prices is quite important to ensure the stable development of economy.

Research has shown that the crude oil price is determined by supply and demand \citep{Hagen2010How,Stevens2007The}. More importantly, the price movement is influenced by some unpredictable extreme events, such as geopolitical conflicts and natural disasters \citep{Bernabe2012A,Ling2015A}. The historical crude oil price reflects the non-linearity, uncertainty, and dynamics, making the accurate forecasting a difficult task. As a result, the uncertain forecasting results are doomed to cause significant uncertainty in the returns of relevant investors and the stable development of the economic system \citep{Zhang2015A}. Thus, it is critical to develop reliable methods for crude oil price forecasting.

Many attempts have been made on forecasting crude oil price, which can be grouped into 2 categories. Traditional statistical methods, such as autoregressive integrated moving average (arima) \citep[e.g.,][]{mohammadi2010international,xiang2013application} and generalized autoregressive conditional heteroskedasticity (garch) \citep{hou2012nonparametric}, have been widely implemented for crude oil price forecasting. On the other hand, machine learning  based methods like support vector machines (SVMs) \citep[e.g.,][]{xie2006new,jun2009oil}, decision trees \citep[e.g.,][]{ekinci2015optimizing,gumus2017crude}, and neural networks \citep[e.g.,][]{movagharnejad2011forecasting,moshiri2006forecasting}, have flourished 
in this area and produced comparable forecasting performance to that of traditional statistical methods. 

With the rapid development of social media, the emergence of user-generated content (UGC) has brought about challenges and opportunities to the area of forecasting. Methods for processing text data have constantly appeared \citep[e.g.,][]{berry2004survey,aggarwal2012mining,Shriharir2015A} and got increasingly mature. Some studies have suggested that the information extracted from the UGC can contribute to the prediction of financial data \citep{Demirer2010The,Kaiser2010The}. Online news is an important part of UGC and contains rich and valuable information which can be utilized to quantify the changes of the public's mood \citep{Serrano2015Sentiment} and the market. 

A series of studies focus on constructing textual features using classic text mining methods and then adopt the combination of textual and non-textual factors for forecasting. \citet{wang2004novel} provide a hybrid AI system framework utilizing the integration of neural networks and rule-based expert systems with text mining. \citet{yu2005rough} investigate a knowledge-based forecasting method, the rough-set-refined text mining (RSTM) approach, for crude oil price tendency forecasting. \citet{li2018text} combine daily WTI futures contract price traded on the New York Mercantile Exchange (NYMEX), US Dollar Index (USDX), and Dow Jones Industrial Average (DJIA) and information such as topics and sentiment extracted from news headlines to forecast crude oil prices, yielding good forecasting performance. Internet searching is also identified as a way of quantifying investor attention and helping forecast crude oil prices \citep{wang2018crude}. \citet{elshendy2018using} incorporate the sentiment of four media platforms (Twitter; Google Trends; Wikipedia; Global Data on Events, Location, Tone database) to forecast the crude oil prices and achieved higher performance.

Our framework of crude oil price forecasting incorporating news text is in line with the work in \citet{li2018text}, where they infer potential topics from news headlines with latent Dirichlet allocation (\emph{LDA}) to forecast crude oil prices. News headlines can be arbitrary, noisy, and ambiguous \citep{shi2018short} as they have only a few words. However, it is worth pointing out that classic topic mode \emph{LDA} is specially designed for accommodating long text \citep{shi2018short}. This mismatch between \emph{LDA} and short news headlines may directly lead to bad effect on the forecasting performance. To tackle the lack of contextual information and further improve forecasting performance, this paper explores to design a novel topic indicator for sparse and short news headlines with semantics-assisted non-negative matrix factorization (\emph{SeaNMF}) \citep{shi2018short}. The widely used \emph{SeaNMF} adapts skip-gram and negative sampling technology to tackle the problem of lack of contextual information and has achieved great success in short text topic modelling \citep{shi2018short}. 

Apart from discovering potential topics from news headlines, \citet{li2018text} also quantify the sentiment of the future market to forecast crude oil prices. Specifically, they construct a relatively static sentiment indicator that simply average the sentiment value of all daily news. This approach is easy to follow, but somehow ignoring the dynamic and complex relationship between the historical and 
current news \citep{xu2014stock}.
To remedy this, we propose a novel dynamic indicator that takes the cumulative and diminishing effect of sentiment into consideration, with the aim at capturing the dynamic information of the changing market.

Another significant characteristic of the work \citep{li2018text} is that they further manually select some exogenous variables such as New York Mercantile Exchange (NYMEX), US Dollar Index (USDX) and Dow Jones Industrial Average (DJIA) to improve the forecasting performance. It is reasonable to choose some non-textual factors to obtain more reliable and accurate forecasts. However, (i) this manual choice of exogenous variables highly depends on professional experts and knowledge, making forecasting results affected by subjective experience to some extent, and more importantly (ii) their work does not systematically examine whether textual features or these exogenous features lead to good predictions.
Thus, there is a need to exploit advanced text mining methods to construct high quality features from short and sparse text and mitigate the importance of manual intervention. Our paper aims to fully amplify the power of text features by utilizing advanced methods without considering extra exogenous variables.


In this paper, we propose a framework for forecasting crude oil price incorporating news headlines. We exploit advanced and appropriate text mining approaches to construct high quality features from sparse and short news headlines, with the aim at tackling the lack of contextual information and extracting as much valuable information as possible. Specifically, this paper constructs a novel topic indicator for future market with \emph{SeaNMF} \citep{shi2018short}. Additionally, a dynamic sentiment indicator taking the cumulative and diminishing effect of market into consideration is proposed. These two marketing indexes are systematically incorporated with \emph{AdaBoost.RT}, yielding better forecasting performance than the benchmarks.

The main contributions of this paper are two-fold:

\begin{itemize}
\tightlist
\item[(1)]Two novel indicators of topic and sentiment specifically for short and sparse news headlines are introduced for crude oil price forecasting, thus forecasting performance is improved.

\item[(2)]Our proposed approach is robust and flexible. In particular to the latter, our framework can be applied to forecast other futures commodities and also yields good forecasting performance.

\end{itemize}

The rest of the paper is organized as follows. Section~\ref{preliminaries} introduces preliminaries for our proposed method. 
Section~\ref{Text-driven crude oil price forecasting} presents the framework of crude oil price forecasting incorporating news text. In Section~\ref{Application to Crude Oil Price Data}, we systematically investigate the forecasting performance of our proposed two market indicators with some benchmarks. Section~\ref{Application to Other Commodities Price Data} applies the proposed method to natural gas and gold prices. Section~\ref{Discussion} gives some discussions and Section~\ref{Conclusion and future work} concludes the article.

\section{Preliminaries}
\label{preliminaries}

\subsection{Text mining related technology}
\subsubsection{GloVe pretrained model for word embedding}
Preprocessing is a fundamental step in text mining, including word tokenization, stop-word filtering and word embedding. The purpose of the first two steps is to transform the text into a collection of words after deleting the unimportant ones. In short, word embedding is a dimension reduction technique that maps high-dimensional words (unstructured information) to low-dimensional numerical vectors (structured information). In other words, word embedding aims to convert documents into mathematical representations as computer-readable input, and thus is an essential work for text analysis problems.\\ 
Two main models: (i) global matrix factorization methods like Latent Semantic Analysis (LSA) \citep{deerwester1990indexing} and (ii) local context windows like skip-gram \citep{mikolov2013efficient} have succeeded in learning word vectors.  However, these methods have some obvious drawbacks. It is poor for LSA to be employed on the word analogy task due to its sub-optimal structure \citep{pennington2014GloVe}. Skip-gram is trained on separate local context windows, making it fail to capture the global information of the corpus  \citep{pennington2014GloVe}.\\
To make up for these drawbacks, an unsupervised learning algorithm for word representation called \emph{GloVe} \citep{pennington2014GloVe} is proposed by Stanford University  and is a new global log-bilinear regression model that aims to combine the advantages of the global matrix and local context window methods.  Especially, \emph{GloVe} explores the training of a word-word co-occurrence matrix instead of the entire sparse matrix. Due to the fact that \emph{GloVe} uses the global and local statistical information of the words to generate a vectorized representation of the language model and words, it is a popular word vector representation in the field of natural language processing.  \\
\emph{GloVe} considers the co-occurrence relationship of words to construct the embedding matrix. We define $X_{ij}$ as the number of times word \emph{j} appears in the context of word \emph{i}. $X_i =\sum_{k}X_{ik} $ is the sum of the number of times any word appears in the context of word \emph{i}. $P_{ij}=P(j|i)=X_{ij}/X_{i}$ is the probability that word \emph{j} appears in the context of word \emph{i}. The co-occurrence probability is defined to calculate the vector representation of word $\tilde{w}_{k}$ when word $w_i$ and $w_j$ are given:
\begin{equation}
F\left(w_{i}, w_{j}, \tilde{w}_{k}\right)=\frac{P_{i k}}{P_{j k}}
.\end{equation}
We expect to maintain the linearity of \emph{F} during the embedding process, so we rewrite \emph{F} as:
\begin{equation}
F\left(w_{i}, w_{j}, \tilde{w}_{k}\right)=F\left(\left(w_{i}-w_{j}\right)^{T} \tilde{w}_{k}\right)=\frac{F\left(w_{i}^{T} \tilde{w}_{k}\right)}{F\left(w_{j}^{T} \tilde{w}_{k}\right)}=\frac{P_{ik}}{P_{jk}}
.\end{equation}
When \emph{F} is an exponential function, this relationship is satisfied, that is $F(x)=exp(x)$, and 
\begin{equation}
\label{fm_log}
w_{i}^{T} \tilde{w}_{k}=\log \left(P_{i k}\right)=\log \left(X_{i k}\right)-\log \left(X_{i}\right)
.\end{equation}
Since $\log(X_{i})$ is a constant term with respect to \emph{k}, it can be written as two bias terms, and formula~\ref{fm_log} is changed to:
\begin{equation}w_{i}^{T} \tilde{w}_{k}+b_{i}+\tilde{b}_{k}=\log \left(X_{i k}\right).\end{equation}
At this time, \emph{w} and \emph{b} form an embedding matrix. 

\subsubsection{{SeaNMF} for short and sparse text topic modelling}
The latent Dirichlet allocation (\emph {LDA}) model is widely used in topic modelling and makes the generative assumption that a document belongs to a certain number of topics \citep{blei2003latent, mazarura2015topic}. However, the \emph {LDA}  model is sensitive and fragile when dealing with the sparse, noisy, and ambiguous of short text. Thus, inferring topics from short and sparse text has become a critical but challenging task \citep[e.g.,][]{chen2011short,jin2011transferring,mazarura2015topic,qiang2017topic}. 


To capture the relationship between a word and its corresponding content in a small window and implement short text topic modelling, \emph{SeaNMF} \citep{shi2018short} employ some key technologies like skip-gram and negative sampling and treat each short text as a window, which can be viewed as word co-occurrence, making it possible to overcome the data sparsity problem. The authors experiment with Tag.News, Yahoo.Ans, and other short text datasets and achieve better results than the \emph {LDA} topic model. A brief description of \emph{SeaNMF} is as follows.

Given a corpus with \emph{N} documents and \emph{M} words, we obtain the word-document matrix \emph{A} and the word-context matrix \emph{S}. $A \in \mathbb{R}_{+}^{M \times N}$ and each column of \emph{A} is the word representation of one document in terms of \emph{M} words. Each element in \emph{S} is the co-occurrence probability of word-context pairs obtained through skip-gram and negative sampling. Our goal is to find lower-rank representations of matrices \emph{A} and \emph{S}: latent matrix \emph{W} of words, latent matrix $W_c$ of context, and latent matrix \emph{H} of document, s.t  $A = WH^{T}$, $S = WW_{c}^{T}$. The relationship among \emph{W}, $W_c$ and \emph{H} is as Fig.~\ref{seanmf}:

\begin{figure}[h!]
  \centering
  \includegraphics[width=11cm,height=6cm]{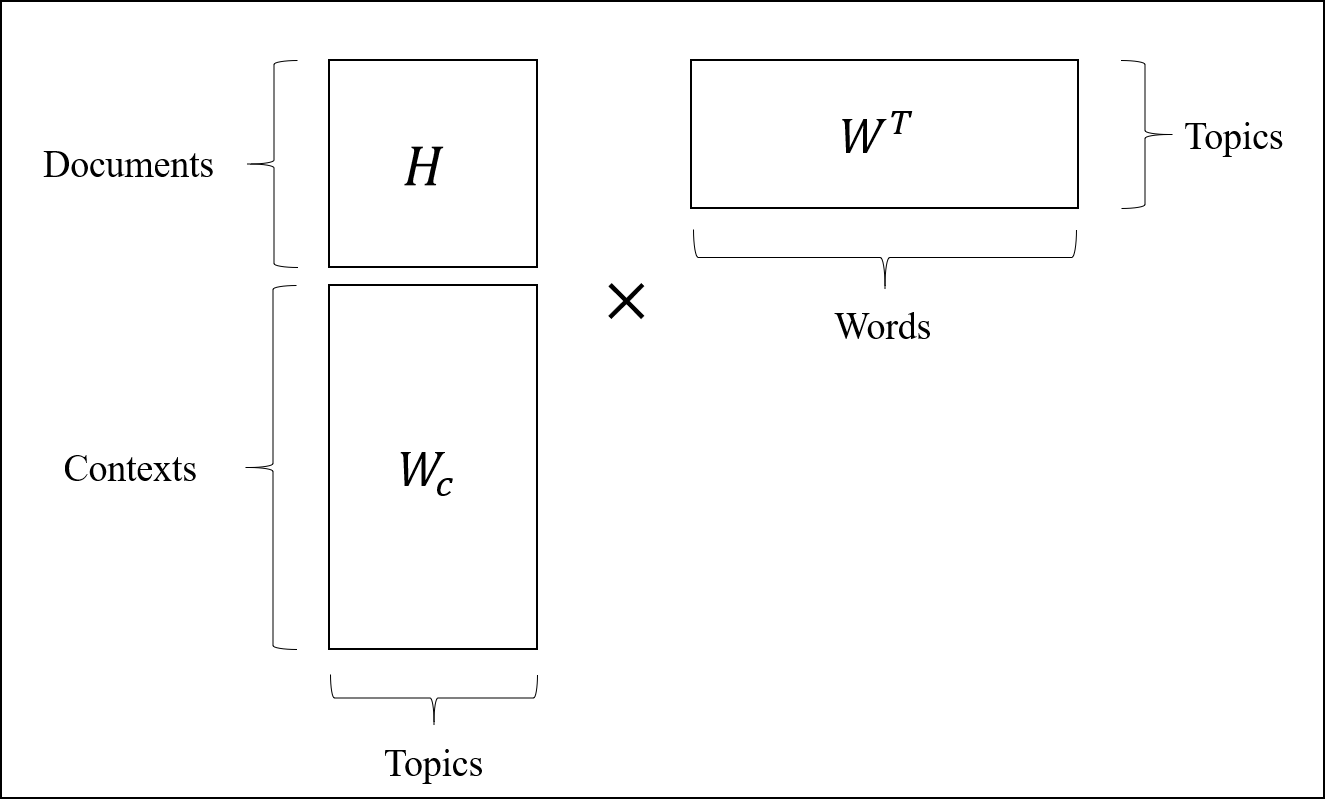}
  \caption{The relationship among \emph{W}, $W_c$ and \emph{H} in \emph{SeaNMF}.}
  \label{seanmf}
\end{figure}

\emph{W}, $W_c$, and \emph{H} are updated in each calculation. For more details, please refer to \citet{shi2018short}. We pay more attention to the matrix \emph{H}, since it contains the weight distribution information of each document on different topics.

\subsection{Time series related technology}
\subsubsection{Order selection for multivariate time series} 
Dependence within and across the series is widely used for time series modelling.
In univariate auto-regression, it is assumed that the current observation of the series is determined by the previous value. While for vector autoregressive (\emph{VAR}), it takes both dependence into consideration and aims to capture the interrelationship among multiple stationary time series when modelling.
The general \emph{VAR} ($p$) is as follows:
 \begin{equation}
  \begin{aligned}
  \mathbf{y}_{t}=\bm{\nu}+A_{1} \mathbf{y}_{t-1}+\cdots+A_{p} \mathbf{y}_{t-p}+\mathbf{u}_{t},
  \end{aligned}
\end{equation}
where $\mathbf{y}_{t}=\left(y_{1 t}, \ldots, y_{K t}\right)^{\prime}$ is a $(K \times 1)$ vector, the $A_{i}$ are fixed $(K \times K)$ coefficient matrices, $\bm{\nu}=\left(\nu_{1}, \ldots, \nu_{K}\right)^{\prime}$ is a fixed $(K \times 1)$ vector of intercept terms allowing for the possibility of a nonzero mean $E\left(\textbf{y}_{t}\right) .$ Finally, $\mathbf{u}_{t}=$ $\left(u_{1 t}, \ldots, u_{K t}\right)^{\prime}$ is a $K$-dimensional white noise process, that is, $E\left(\mathbf{u}_{t}\right)=\mathbf{0},  E\left(\mathbf{u}_{t} \mathbf{u}_{t}^{\prime}\right)=\Sigma_{u}$ and $E\left(\mathbf{u}_{t} \mathbf{u}_{s}^{\prime}\right)=0$ for $s \neq t .$ 

Combined with probabilistic information criteria, such as \emph{AIC}, \emph{SIC}, \emph{HQ}, etc., the lag $p$ of each time series can be found \citep{lutkepohl2005new}. In short, \emph{AIC} is suitable for small samples, and \emph{SIC} performs well in large samples, according to \citet{ivanov2005practitioner}. So we choose the \emph{SIC} criteria to help find the optimal lag in this paper.
\begin{equation}
SIC(p)=\ln |\bar{\Sigma}(p)|+\frac{\ln N}{N}\left(K^{2} p\right)
,\end{equation}

Where \emph{K} is the dimension of  the \emph{VAR}, and \emph{N} is the sample size. $\bar{\Sigma}(p)$ is the quasi-maximum likelihood estimate of the innovation covariance matrix $\Sigma(p)$. We aim to choose a lag \emph{p} that minimizes the value of the criterion function.

\subsubsection{Time series regression}
\begin{figure}[h!]
  \centering
  \includegraphics[width=16cm,height=9cm]{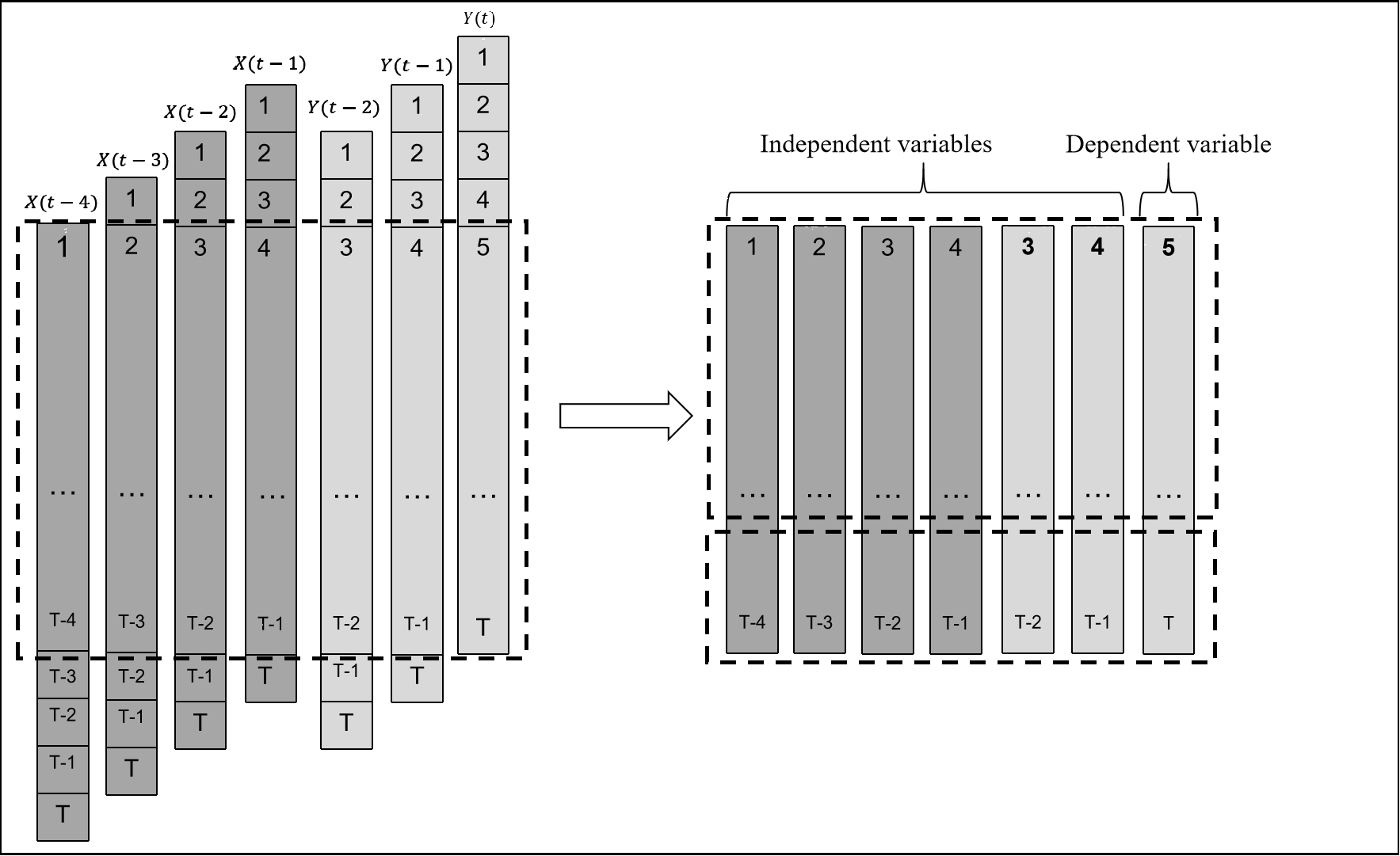}
  \caption{A toy example illustrates how to transform the forecasting of multivariate time series into a regression problem. In this case of daily data, the current observations $Y_t$ can be predicted by the observations of the past 4 and 2 days of \emph{X} and \emph{Y} respectively. Technically, when predicting $Y_t$, we use the observed values of $Y_t$ moving back by two days and the $X_t$ moving back by 4 days as features.}
  \label{fore2reg}
\end{figure}

Time series regression refers that a target variable can be forecast by some regressors.
One common method for forecasting multivariate time series is to convert the forecasting problem into a regression problem. 
We take a simple example to illustrate this method. Given an endogenous variable \emph{Y} with a lag of 2 and an exogenous variable \emph{X} with a lag of 4, we aim to use these lag values to predict \emph{Y}. First, we obtain 4 and 2 copies of \emph{X} and \emph{Y} respectively. Then we shift the copies of \emph{X} and \emph{Y} as shown in the left part of Figure~\ref{fore2reg}, remove the rows where the null values exist, and get the data set of the regression model. Finally, the 2 lags of \emph{Y} are also included in the independent variables, and the regression equation of the independent variable \emph{Y} can be written as formula~\ref{fm_example} where ${Y}_{t}$ can be predicted by these lagged values

\begin{equation}
\label{fm_example}
\hat{Y}_{t}=f(Y_{t-1}, Y_{t-2}, X_{t-1}, X_{t-2}, X_{t-3}, X_{t-4})
.\end{equation}

\subsection{Machine learning related technology}

\subsubsection{RFE for feature selection}
Large volume features may bring redundant noise and result in worse forecasting performance. Feature selection aims to obtain the most relevant set from all the features and as a result reduces the computation time and complexity.

Recursive Feature Elimination (\emph{RFE}) is a common feature selection algorithm. \emph{RFE} is a wrapper included core functions. Given the number of features, the core functions are fitted to rank the features according to their importance. After removing the least important feature, the model is refitted. The process repeats until the number of features we specify are retained \citep{guyon2002gene}. Given a data set with \emph{k} features, and set \emph{F} containing all features initially. The specific steps of \emph{RFE} are as follows:

\textbf{Step 1.} Repeat for $\emph{p} = 1,2,...,k$: \\
\textbf{Step 2.}	\hspace*{0.6cm} \% Do the \emph{RFE} procedure.\\
				\hspace*{2.4cm} Repeat for $\emph{i} =1,2,...,\emph{k} - \emph{p}$:\\
					\hspace*{3.0cm} Fit core function with \emph{F};\\
					\hspace*{3.0cm} Rank \emph{F} according to the feature importance;\\
					\hspace*{3.0cm} $f^{*} \leftarrow$ the least important feature in \emph{F};\\
					\hspace*{3.0cm} $\emph{F} \leftarrow \emph{F} - f^{*}$;\\
				\hspace*{2.4cm}\%  \emph{p} important features remain in \emph{F} after this step.\\
\textbf{Step 3.}	\hspace*{0.6cm} Compute \emph{rmse}, \emph{mae}, \emph{mape} for model with \emph{F}.\\
\textbf{Step 4.} Choose the model corresponding to the minimum mean of the three indicators.

The formulas of \emph{rmse}, \emph{mae}, and \emph{mape} are as follows:
\begin{equation}
\label{fm_rmse}
\emph{rmse} = \sqrt{\frac{1}{n}\sum_{i=1}^{n}(\hat{y_i}-y_i)^2}
,\end{equation}

\begin{equation}
\label{fm_mae}
\emph{mae} = \frac{1}{n}\sum_{i=1}^{n}\left\vert \hat{y_i}-y_i \right\vert
,\end{equation}

\begin{equation}
\label{fm_mape}
\emph{mape} = \frac{100\%}{n}\sum_{i=1}^{n}\left\vert \frac{\hat{y_i}-y_i}{y_i} \right\vert
,\end{equation}

where $\hat y_i$ is the predicted value and $y_i$ is the true value.

\subsubsection{AdaBoost.RT for regression problem}
Two main machine learning technologies bagging and boosting have achieved great success in improving the accuracy through the combination of the predictions of multiple learners \citep{solomatine2004AdaBoost}.  \emph{AdaBoost.RT} belongs to the family of boosting and is an ensemble method for regression.
The obvious advantage of \emph{AdaBoost.RT} from the other methods is the relative error loss functions, making it possible to give enough attention to the examples with lower accuracy \citep{solomatine2004AdaBoost}. 
\\
\emph{AdaBoost} was originally designed as a classification algorithm, and \citet{solomatine2004AdaBoost} propose \emph{AdaBoost.RT} for the regression problem based on \emph{AdaBoost}. \emph{AdaBoost.RT} combines several weak learners to form a strong learner, which can output the  results through adjustment of thresholds and multiple rounds of iterative calculation. Given the features \emph{X} and dependent variable \emph{Y} of the data set, we implement \emph{AdaBoost.RT} through the following steps:\\
\textbf{Step 1.} Initialize \emph{T} weak learners, each with a weight of $1/T$. Thus the weight distribution of these weak learners is $D_t = (1/T, 1/T, ... ,1/T)$. The maximum number of iterations is set to \emph{N}.\\
\textbf{Step 2.}	Repeat for $i = 1, 2, ..., N$: \\
	\hspace*{1.8cm} Fit regression equation $f_t(X) \rightarrow Y$ for each weak learner;\\
	\hspace*{1.8cm} Calculate error rate between $f_t(X)$ and $Y$; \\
	\hspace*{1.8cm} Update $D_t$ according to the error rate;\\
\textbf{Step 3.}	$F(X) \leftarrow \sum_{t} D_t \times {weak learner_t}$

\section{Text-driven crude oil price forecasting}
\label{Text-driven crude oil price forecasting}
The purpose of this study is to establish a time series forecasting framework incorporating text features. Topic and sentiment information can be extracted from a large number of futures-related news headlines through text mining technology. Then, the text-related features can be used for exogenous variables to make predictions. The specific implementation process is shown in Figure~\ref{framework-of-forecasting}. It is necessary to answer the following two questions:
\begin{itemize}
\tightlist
\item[(1)] \textbf{Why headlines instead of full text?} The news headlines is a summary of the full text and can be considered to contain most of the essential information. Also, our work is in line with \citep{li2018text}, to be consistent with their work, we choose news headlines for extracting topic and sentiment information.
\item[(2)] \textbf{Why futures news instead of crude oil news?} There are two reasons for this choice. First, we tried to collect crude oil news but only obtained approximately 2,000. The use of futures news has expanded the text dataset approximately ten times. Second, relevant studies have proven that there are complex correlations among futures prices such as gold, natural gas, and crude oil prices. \citet{sujit2011study} argue that fluctuations in gold prices will affect the WTI index. 
Different countries' dependence on crude oil (import or export) will affect their currency exchange rate and then affect people's purchasing power for gold. In the gold market, if the supply-demand relationship changes, then the prices will change accordingly. \citet{villar2006relationship} observe that a 1-month temporary shock to the WTI of 20 percent has a 5-percent contemporaneous impact on natural gas prices.
\end{itemize}

\subsection{The construction of daily topic intensity for the futures market}

Following the instruction of \emph{SeaNMF} (\url{https://github.com/tshi04/SeaNMF}), we can obtain the topic weight distribution of each headline, from which we calculate the probability of each headline belongs to each topic. To select the number of topics, the \emph{pointwise mutual information (PMI)} score is calculated \citep{quan2015short}. Given a set of topic numbers, \emph{PMI} evaluates the effectiveness of the model and chooses the optimal number of topics.
Due to the fact that the media publishes a lot of news every day, we calculate the average weight of news as the topic intensity of the day. The \emph{topic intensity index} of the \emph{t-th} day is defined as follows:
\begin{equation}
\emph{TI}_\emph{it}=\frac{1}{N_{t}}\sum_{\emph j=1}^{\emph n}\emph {DT}_\emph{ij}
,\end{equation}
where $N_{t}$ is the number of news in one day, $\emph{TI}_\emph{it}$ is the \emph{i-th topic intensity index} of the \emph{t-th} day;
$\emph {DT}_\emph{ij}$ is the weight of \emph{j-th} news of \emph{i-th} topic in \emph{t-th} day.

\subsection{The construction of daily sentiment intensity considering the effect of exponential decay}
\label{sentiment-anlysis}
With the rapid development of social media, people have more channels to publish and read text messages, which contain different sentiments and attitudes of the public. Taking futures-related news as an example, the positive and negative sentiments within the news often affect people's judgment on the changing futures market, which can be reflected in the fluctuation of the futures prices. 

Sentiment analysis is a key technology for text mining. It adopts computer linguistic knowledge to identify, extract, and quantify sentiment information in the text. \emph{TextBlob} (\url{https://textblob.readthedocs.io/en/dev/}), as a python library that can handle a variety of complex NLP problems, is practical to calculate the sentiment score of one piece of news text. \emph{TextBlob} has a huge built-in dictionary. When calculating the sentiment polarity of a sentence, it traverses all the words in the sentence and averages them through the labels of the dictionary to calculate the sentiment score. \emph{TextBlob} is quite simple to use, and can effectively deal with the modifiers and negative words in the sentence, thus it's an effective tool for many studies 
\citep[e.g.][]{kaur2020twitter,kunal2018textual,saha2017proposed}. The sentiment scores range from -1 to 1, and the smaller the value is, the more negative, and vice versa.

By simply averaging the sentiment scores of all news headlines in one day, we can obtain the sentiment intensity of this day.
\begin{equation}
\emph S\emph V_t=\frac{1}{N_{t}}\sum_{\emph i=1}^{\emph N_t}\emph P\emph V_{it}  ,
\end{equation}
where $\emph P\emph V_{it}$ represents the sentiment value of the $\emph {i-th}$ news items on the $\emph {t-th}$ day, and $\emph N_t$ is the number of news items published on the $\emph {t-th}$ day. The $\emph{SV}_t$ refers to the average sentiment intensity of the $\emph {t-th}$ day.

However, the impact of news on people's sentiment is often continuous in the actual futures market. That is, on a specific day, public sentiment is the result of the combination of the news on current and that in the previous days, except that the current news is more influential. Given this complex situation, it is assumed that the impact of news on public sentiment is exponentially attenuated. Considering the sentiment continuity, we design a \emph{sentiment index (SI)} $\emph e^{-\frac{m}{7}}$ considering the effect of exponential decay
inspired by the work \citep{xu2014stock}, which is more in line with the actual situation of news impact. It is assumed that news has the strongest impact on crude oil prices for the next seven days. $\emph m$ represents the number of days after the news release. For instance, on the release day, $\emph m = 0$, $\emph {SI}=\emph e^{-\frac{0}{7}}=1$ ; when $\emph m=1,     \emph {SI}=\emph e^{-\frac{1}{7}}=86.69\%$, the following \emph{SIs} are $75.15\%,  65.14\%,...$.

The $\emph{sentiment intensity}$ on the $\emph{t-th}$ day is the sum of the $\emph{SV}$ on the $\emph{t-th}$ day and the  \emph{SVs} in the previous days.

\begin{equation}
\label{sentiment intensity}
\emph{SI}_t = \sum_{\emph i = 1}^{\emph t-1}\emph e^{-\frac{\emph{t-i}}{7}}\emph{SV}_\emph i+\emph{SV}_\emph t
,\end{equation}
$\emph{SI}_t$ is the \emph{sentiment intensity} of the \emph{t-th} day. $e^{-\frac{\emph{t-i}}{7}}\emph{SV}_\emph i$ is the sentiment impact of the \emph{i-th} day on the \emph{t-th} day.

The \emph{sentiment intensity} we designed has the following key innovations:

\begin{itemize}
\tightlist
\item[(1)] The cumulative effect of sentiment is considered. In addition to the sentiment calculation of news on the release day, the current sentiment will be affected by the previous news, which is more in line with the actual situation;

\item[(2)] The exponential diminishing effect of sentiment is considered. With the continuous release of news, people will gradually forget the past information, and as a result the influence of the early news will be weakened. We aim to capture this diminishing effect by the means of exponential decay.
\end{itemize}

\subsection{The general framework of crude oil price forecasting incorporating news headlines}

\begin{figure}[h!]
  \centering
  \includegraphics[width=17cm,height=8cm]{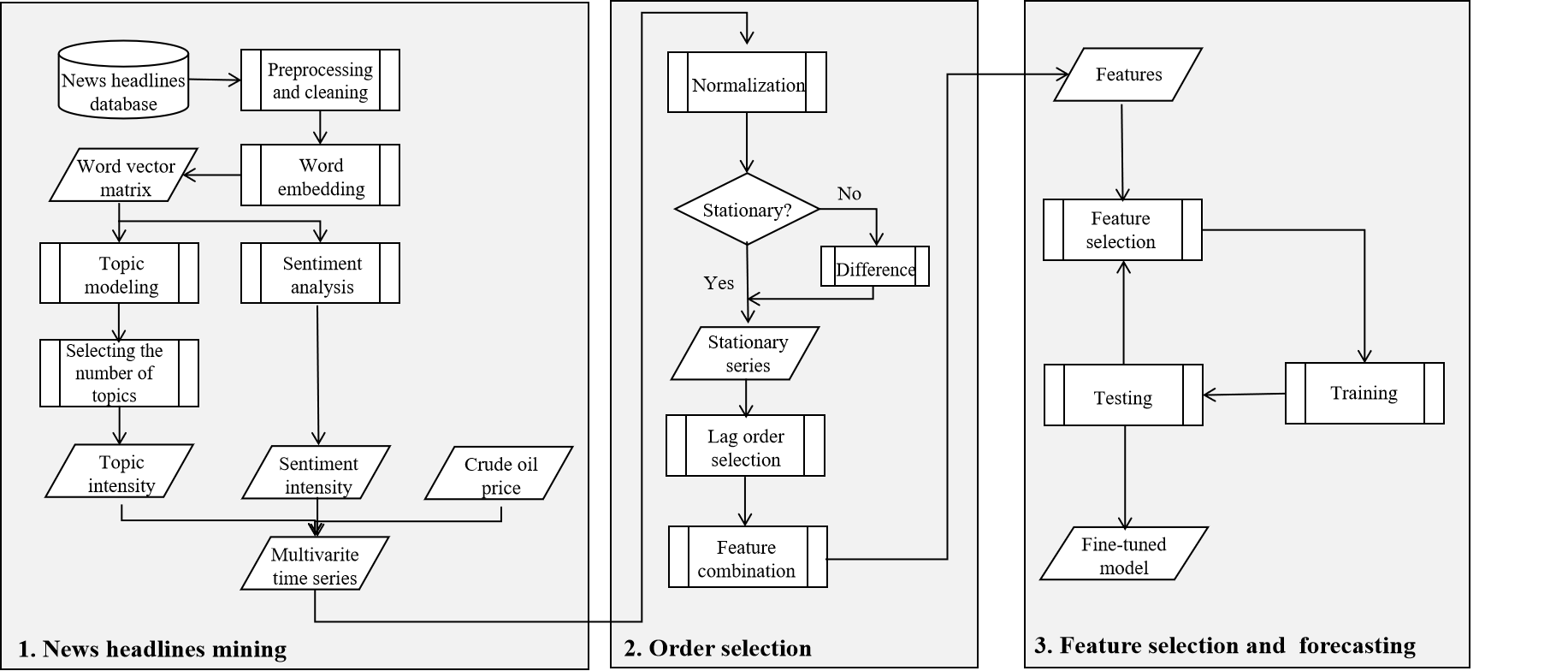}
\caption{The framework of crude oil price forecasting. }
  \label{framework-of-forecasting}
\end{figure}

Fig.~\ref{framework-of-forecasting} shows the proposed forecasting framework. Although our work is line with  \citet{li2018text}, it should be emphasized that this research focuses on point forecasting rather than trend forecasting. We particularly focus on the design and employment of appropriate methods for modelling short and sparse news headlines. 

The crude oil forecasting incorporating news text includes three main parts: 
\begin{enumerate}
\tightlist
\item \textbf{News headlines mining}: the news headlines are first preprocessed, including word segmentation, stop words filtering, stem extraction, etc. Then we use \emph{GloVe} to do word embedding for the clean texts and get the word vector matrix. Subsequently, topic modelling and sentiment analysis are employed to calculate the topic intensity and sentiment intensity of the futures market.
\item  \textbf{Order selection}: we carry out the first-order difference processing for non-stationary time series. We respectively model the interrelationship between each exogenous series with crude oil price series with \emph{VAR} and obtain the optimal lag. Then we covert multivariate time series forecasting into a regression problem based on these lag value.
\item  \textbf{Feature selection and forecasting}: we use \emph{RFE} to select the most relevant features when constructing the forecasting model. By building a variety of models and comparing \emph{rmse}, \emph{mae}, and \emph{mape}, we finally choose the model that performs best.
\end{enumerate}

To compare the proposed forecasting framework with \citet{li2018text}, Table~\ref{we and Li} illustrates the similarities and differences of them. We originally plan to use the the same range of data with \citet{li2018text}, but we cannot have access to the data as \href{https://www.investing.com/}{\emph{Investing.com}}  deleted earlier news. In the following experiments, to be fair and comparable, we reproduce their method  \citep{li2018text} and apply it to our data as one of our benchmarks.



\begin{table*}[!htbp]  
\footnotesize
	\centering	
	\caption{Comparisons between our forecasting framework and \citet{li2018text}.}  
	\label{we and Li}
        \resizebox{460pt}{110mm}{
	\begin{tabularx}{\textwidth}{lXX}
		\toprule
		& \citet{li2018text} & Our forecasting framework \\
                \midrule
                Research objective & Trend forecasting & Point forecasting \\
                \midrule
                \multirow{3}{*}{Date range} & Total range: 2009.9.15-2014.7.20 &Total range: 2011.3.29-2019.3.22\\
                                                               & Training range: 2012.4.18-2013.10.7 & Training range: 2011.3.29-2016.7.22\\
								& Test range: 2013.10.8-2014.7.20 & Test range: 2016.7.23-2019.3.22\\
		\midrule
		Data retrieval & Crude oil price series; Crude oil related news headlines;  Other data from financial market & Crude oil price series; Futures related news headlines\\
		\midrule
		\multirow{3}{*}{Data preprocessing} & 1. Tokenization and stop-words filtering & 1. Tokenization and stop-words filtering\\
									      & 2. Word embedding with Term Frequency –
Inverse Document Frequency (TF-IDF) & 2. Word embedding with \emph{GloVe} \\
									      & 3. Hodrick-Prescott (HP) smoothing for trend forecasting & 3. Without HP smoothing \\
		\midrule
		\multirow{4}{*}{News headlines mining} & 1. CNN is used to predict the next day's oil price movement. & 1. We delete this part for the bad classification accuracy of CNN. \\
									            & 2. Discrete and static sentiment of polarity and subjectivity is calculated by \emph{TextBlob}. & 2. We design a continuous and dynamic sentiment intensity based on an exponential model (see Formula~\ref{sentiment intensity}). \\
										   & 3. Latent topics were discovered by \emph{LDA} from short and sparse news headlines and Dynamic Topic Model (DTM), but \emph{DTM} is worse for its stable topics. & 3. We employ \emph{SeaNMF}, a better topic model for short and sparse texts, with the aim at tackling the lack of contextual information.\\
										   & 4. KL divergence is used to choose the number of topics. & 4. Pointwise Mutual Information (PMI) score was used in our framework.\\
		\midrule
		Lag selection & \emph{VAR} model & \emph{VAR} model \\
		\midrule
		Feature selection & \emph{RFE} model & \emph{RFE} model \\
		\midrule
		Price forecasting & Random forest; Support vector regression; Linear regression & Random forest; Support vector regression; arima and arimax; AdaBoost.RT model \\
		\midrule
		Evaluation & mae; rmse & rmse; mae; mape; Multi-step forecasting, DM tests, the applications to gold and natural gas support  the validity and generalization of  our framework.\\

		\bottomrule
	\end{tabularx}
	      }
\end{table*}

\section{Application to crude oil price data}
\label{Application to Crude Oil Price Data}
\subsection{Data collection and description}
\label{Data collection}
\href{https://www.investing.com/}{\emph{Investing.com}} is a world-renowned financial website that provides real-time information and news about thousands of financial investment products, including global stocks, foreign exchange, futures, bonds, funds, and digital currency, as well as a variety of investment tools. We collected 28,220 news headlines through the futures news column on \href{https://www.investing.com/}{\emph{Investing.com}} as the text data of this study. \\
We collected oil price daily data from March 29, 2011, to March 22, 2019, on \href{https://fred.stlouisfed.org/series/DCOILWTICO}{\emph{FRED Economic Data}}, and the news collected also covered this period. The selected base oil is West Texas Intermediate (WTI) crude oil, which is a common type in North America. WTI has become the benchmark of global crude oil pricing due to US military and economic capabilities in the world.
\subsection{Experimental design}
\begin{figure}[htbp]
  \centering
  \includegraphics[width=12cm,height=6cm]{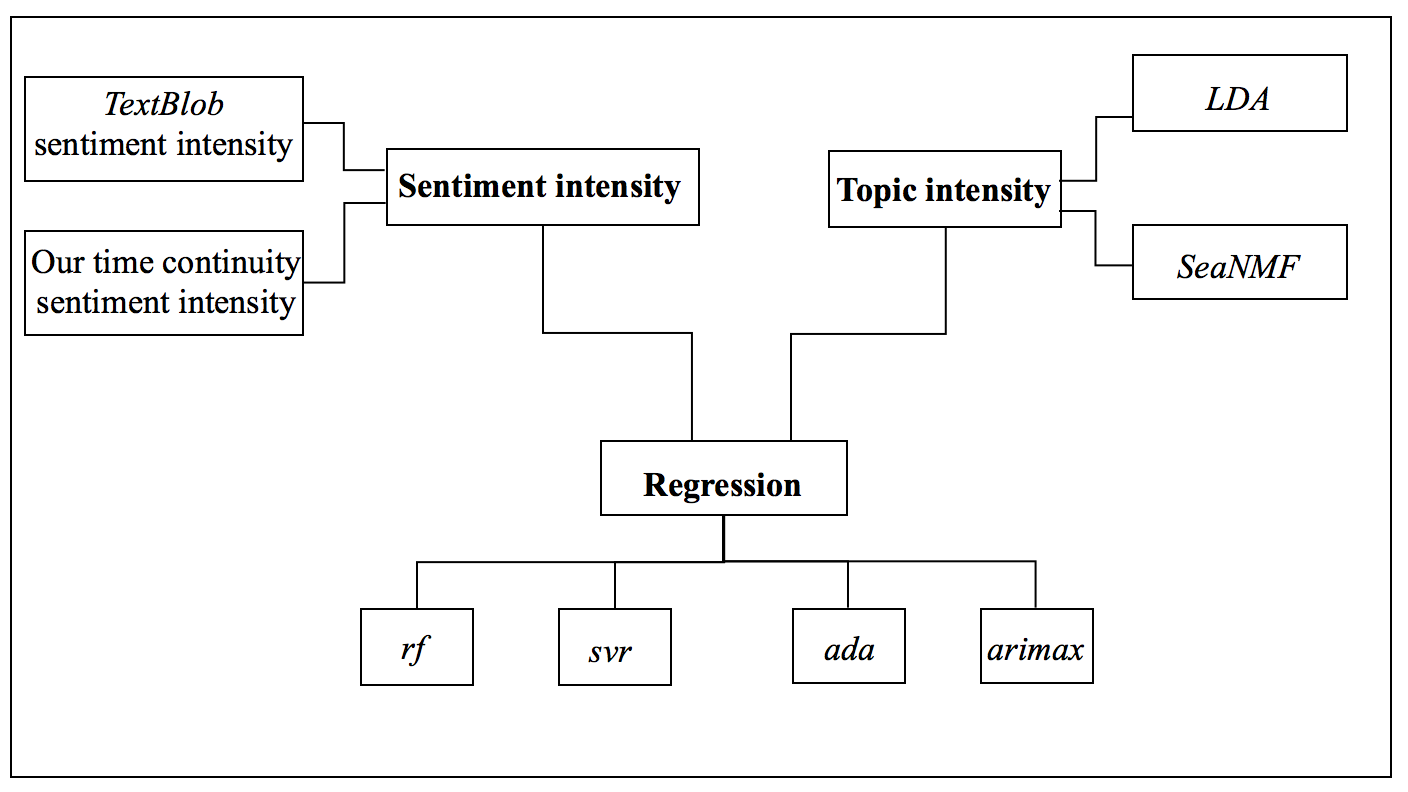}
\caption{Experimental design.}
  \label{experiment}
\end{figure}

In order to systematically and comprehensively verify the superiority of our proposed topic and sentiment indicators, Figure~\ref{experiment} presents the framework of our experimental design. Due to time and energy limitation, we just choose some classic and common used methods regrading to the sentiment analysis, topic modelling, and regression models as our candidate sub-models.
Thus, the experimental process includes the comparison of the multiple models based on these sub-models.
\begin{itemize}
\item[(1)] For the sentiment analysis, we choose a widely used sentiment index integrated in \emph{TextBlob}. We aim to compare its ability to forecast crude oil price with our novel sentiment intensity. 
\item[(2)] In the part of topic modelling, \emph{LDA} and \emph{SeaNMF} are compared for short and spare news headlines. 
\item[(3)] The regression models have been introduced separately in Table~\ref{methods_compare}. 
\end{itemize}

\begin{table}
  \caption{\label{methods_compare} Description of the other five forecasting methods}
\centering
\resizebox{\linewidth}{!}{
\begin{tabular}[t]{lp{40em}}
  \toprule
  Method & Description                                                                                                                                                                                                                                           \\
  \midrule
     rf   & \emph{rf} is a bagging technology that trains multiple decision trees in parallel and outputs the average prediction results of these trees \citep{liaw2002classification}.                                                                                                        \\
  svr  &  The purpose of \emph{svr} is to find the optimal decision boundary so that the data points are closest to the hyperplane or the support vectors are all within the boundaries \citep{drucker1997support}.                                                                              
 \\
  arima  & \emph{arima} is a well-known time series forecasting model. It is a linear equation whose predictors include the lags of the dependent variable and the lags of the forecasting errors \citep{contreras2003arima}.                                                                                                                                                                                                                                                                                                                                                           \\
arimax & \emph{arima} is suitable for univariate time series forecasting, while \emph{arimax} performs well on multivariate analysis \citep{hyndman2010arimax}.
\\
  \bottomrule
\end{tabular}}
\end{table}

\subsection{LDA versus SeaNMF topic analysis for short and sparse news headlines}
\begin{figure}[h!]
  \centering
  \includegraphics[width=12cm,height=8cm]{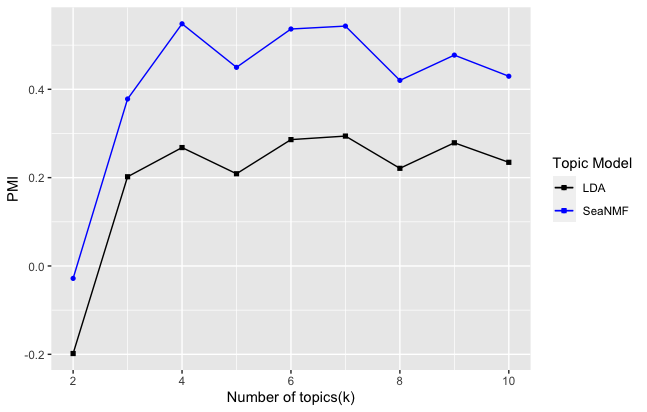}
  \caption{Comparison of the \emph{SeaNMF} and the \emph{LDA} for short and sparse news headlines.}
  \label{PMI}
\end{figure}

The \emph{PMI} score is used to compare the performance of \emph{LDA} and \emph{SeaNMF} topic models. The higher the \emph {PMI} score, the better the model performance. We set \emph k from 2 to 10 to calculate the \emph{PMI} scores in turn. The blue line in Figure~\ref{PMI} represents the \emph{PMI} value of the \emph{SeaNMF}, and the black line represents the \emph{PMI} value of the \emph{LDA}. It can be seen from the figure that the \emph{PMI} value of \emph{SeaNMF} is generally higher than that of \emph{LDA} and relatively stable. This shows that \emph{SeaNMF} is better than \emph{LDA} in extracting topics from news headlines. When $k=4$, the \emph{PMI} value of \emph{SeaNMF} is the highest, indicating that the model works best when the number of topics is 4. As the number of topics increases, the \emph{PMI} value of \emph{LDA} shows a decreasing trend and fluctuates greatly. Therefore, we will no longer consider using \emph{LDA} to extract topics in following experiments.

We select the top 10 keywords from each topic of \emph{SeaNMF}, as shown in Table~\ref{keywords}. From the keywords, we can see that the \emph{SeaNMF} model can indeed extract different and meaningful topics from the text. The bold font shows that the four topics can be approximately summarized as crude oil, gold, natural gas, and new energy respectively.

\begin{table}[htbp]
\footnotesize   
  \centering
  \caption{Top 10 keywords of 4 topics for \emph{SeaNMF} model}
  \label{keywords}
  \resizebox{340pt}{18mm}{
    \begin{tabular}{llllllll}
      \toprule
      Topic        &Keywords\\
      \midrule
      1 & \textbf{oil crude} u.s prices data supply opec asia ahead gains \\
      2 & \textbf{gold} prices fed asia dollar u.s data ahead gains higher \\
      3 & futures \textbf{gas natural} u.s weekly outlook data low weather supply \\
      4 & exclusive says \textbf{energy new sources} trump billion coal pipeline saudi \\
      \bottomrule
    \end{tabular}
  }
\end{table} 

\subsection{Order selection}
\begin{figure}[htbp]
  \centering
  \includegraphics[width=16cm,height=10cm]{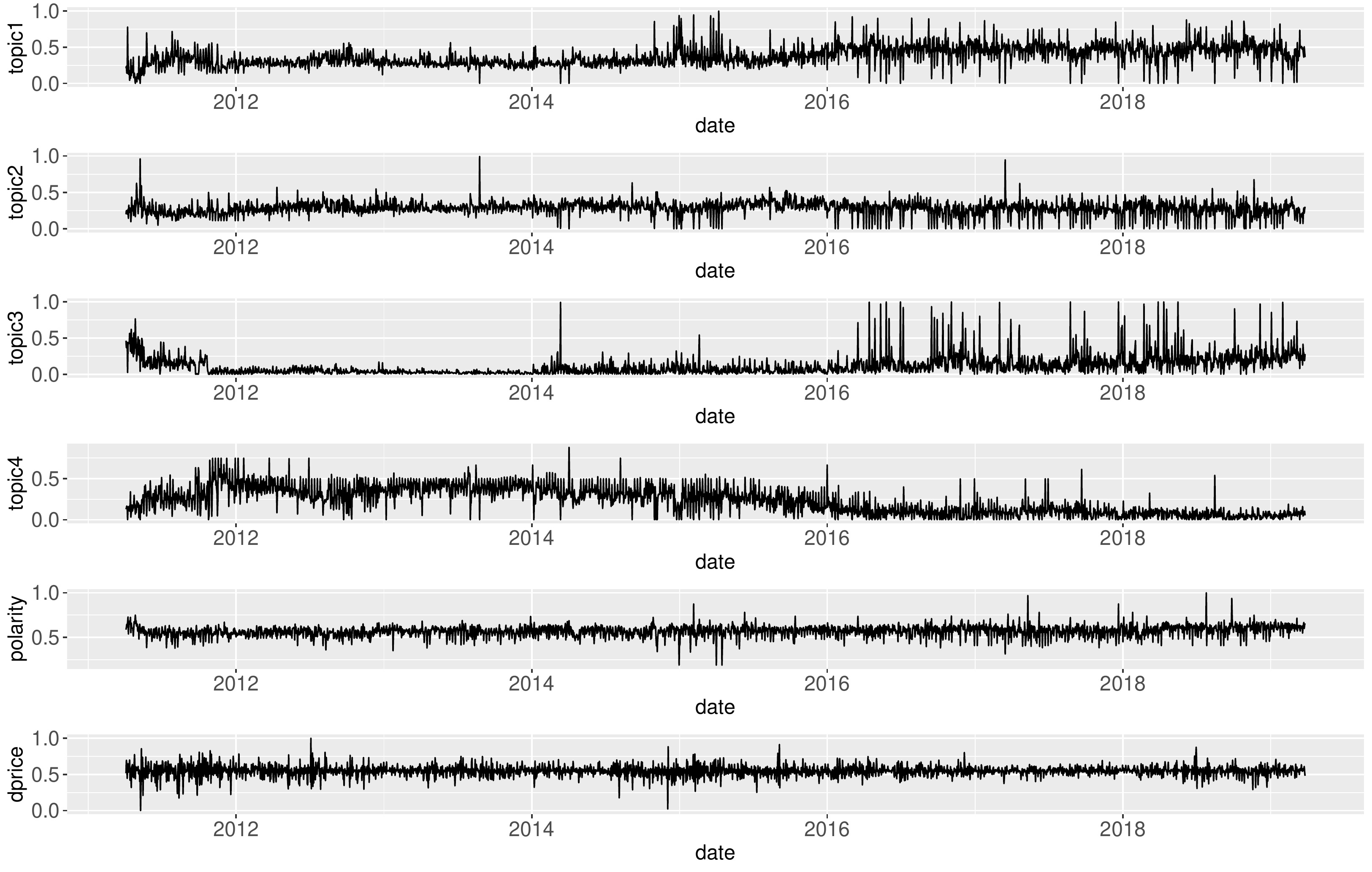}
\caption{Time series of crude oil price and text features. }
  \label{6 time series}
\end{figure}

After calculating the topic and sentiment intensity, we obtain six time series, including topic 1 to topic 4, sentiment index, and crude oil prices. Then, we respectively model the interrelationship between each exogenous series with crude oil price series with \emph{VAR} and obtain the optimal lag. The results are shown in Table ~\ref{Lag-order}, in which \emph{dprice} means that the original price series is non-stationary, and changes to stationary after the first order difference. \emph{polarity} is the \emph{sentiment intensity}. All the series are shown in Figure~\ref{6 time series}, and the description of them are listed in Table~\ref{Description of 6 time series}. We can write the regression equation in the following formula where crude prices can be predicted by these lag values.
\begin{equation}
\begin{aligned}
\begin{split}
\hat{dprice_t} = &f(dprice_{t-1},  dprice_{t-2},  dprice_{t-3},  topic 1_{t-1},...,  topic 1_{t-7}, \\
&topic 2_{t-1},..., topic 2_{t-7}, topic 3_{t-1},..., topic 3_{t-7}, \\
&topic 4_{t-1},..., topic 4_{t-7}, polarity_{t-1},..., polarity_{t-7}).
\end{split}
\end{aligned}
\end{equation}

\begin{table}[htbp]
\footnotesize   
  \centering
  \caption{Optimal lag of 6 time series related to crude oil}
  \label{Lag-order}
  \resizebox{340pt}{11mm}{
    \begin{tabular}{llllllll}
      \toprule
      Time Series  & topic 1 &  topic 2 & topic 3 &topic 4 &polarity &dprice \\
      \midrule
      SIC & -13.0595 & -13.4605 & -12.9977 & -13.4579 & -14.3271 & -8.5941 \\
      Lag  & 7 & 7 & 7 & 7 & 7 &3\\
      \bottomrule
    \end{tabular}
  }
\end{table}

\begin{table}[htbp]
\footnotesize   
  \centering
  \caption{Description of 6 time series}
  \label{Description of 6 time series}
  \resizebox{340pt}{16mm}{
    \begin{tabular}{llllllll}
      \toprule
        & topic 1 &  topic 2 & topic 3 &topic 4 &polarity &dprice \\
      \midrule
      mean & 0.3743 & 0.2835 & 0.1157 & 0.2264 & 0.5077 &0.5477 \\
      median & 0.3559 & 0.2885 & 0.0723 & 0.2175 & 0.5763 &0.5505\\
      std & 0.1330 &0.0925 & 0.1341 & 0.1587 & 0.0601 &0.0748\\
      \bottomrule
    \end{tabular}
  }
\end{table}

\subsection{Feature selection and forecasting}
\label{Feature selection and forecasting}
\begin{figure}[htbp]
  \centering
  \includegraphics[width=9cm,height=2cm]{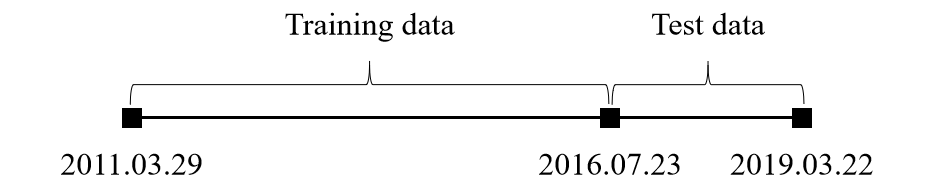}
\caption{The data in the training set is from March 29, 2011 to July 22, 2016, and the data in the test set is from July 23, 2016, to March 22, 2019. }
  \label{split}
\end{figure}

The time series is divided into a training set and a test set as shown in Figure~\ref{split}. The performance of forecasts up to 3 days ahead is evaluated over the test data.

After obtaining the lag values of the time series, it is intuitive to regard them as independent variables and oil prices as dependent variables to train the regression model. We use \emph{RFE} to select features and random forest regression (\emph{rf}), support vector regression (\emph{svr}), autoregressive integrated moving average (\emph{arima}), autoregressive integrated moving average with explanatory variable (\emph{arimax}), the method from \citet{li2018text} (\emph{svr-Li}) and AdaBoost.RT (\emph{ada}) to fit the crude oil prices and complete forecasting on the test set.  \emph{svr-Li} is a forecasting model based on \emph{svr} that combines multi-source text and financial features \citep{li2018text}.
At the same time, the \emph{rmse}, \emph{mae}, \emph{mape} between the model with and without text features are compared. Some brief introductions have been listed in Table~\ref{methods_compare}.

\begin{table}[htbp]  
\footnotesize
	\centering
	\caption{Forecasting results of multiple methods for crude oil price over h=1, 2 and 3.}  
	\label{Comparison of Models for Crude Oil Price Forecasting}
      \resizebox{460pt}{42mm}{
	\begin{tabular}{ccc ccc ccc ccc}
		\toprule
		&\multirow{2}{*}{Model} & Number of& \multicolumn{3}{c}{h=1} &  \multicolumn{3}{c}{h=2} & \multicolumn{3}{c}{h=3} \\
		\cmidrule(lr){4-6} \cmidrule(lr){ 7-9} \cmidrule(lr){10-12} 
		 & & features & \emph{rmse} & \emph{mae} & \emph{mape} & \emph{rmse} &\emph{mae} & \emph{mape}  & \emph{rmse} &\emph{mae} & \emph{mape} \\
		\midrule
		\multirow{3}{*}{rf} & no text & 3  & 0.0728 & 0.0545 & 0.1018 & 0.0742 & 0.0561 & 0.1042 & 0.0744 & 0.0564 & 0.1052 \\
		& textblob & 36      & 0.0681 & 0.0514 & 0.0971 & 0.0691 & 0.0526 & 0.0989 & 0.0676 & 0.0504 & 0.0947 \\
                & our method & 27  & 0.0683 & 0.0520 & 0.0968 & 0.0714 & 0.0542 & 0.1009 & 0.0720 & 0.0531 & 0.0985   \\
		\midrule
		\multirow{4}{*}{svr} &svr-Li  & 28 & 0.0646 & 0.0475 & 0.0885 & 0.0602 & 0.0444& 0.0827 & 0.0603 & 0.0441 & 0.0821\\
                &no text & 2 & 0.1111 & 0.1001 & 0.1765 & 0.1116 & 0.0999 & 0.1761 & 0.1116 & 0.1002 & 0.1766\\
		&textblob &14    & 0.0586 & 0.0427 & 0.0799 & 0.0597 & 0.0442 & 0.0827 & 0.0593 &0.0428 &0.0802 \\ 
		& our method &17   & 0.0584 & 0.0428 & 0.0801 & 0.0596  & 0.0439 & 0.0819 & 0.0593 & 0.0428 & 0.0801 \\
		\midrule
                \multirow{3}{*}{arima \& arimax} & no text &  --   &0.0565 & \bf 0.0394 & 0.0751 & 0.0566 & \bf 0.0394 & 0.0753 & 0.0566 & \bf 0.0394 & 0.0754\\
		&textblob& 23 & 0.0568 & 0.0401 & 0.0757 & 0.0568 & 0.0404 & 0.0759 & 0.0573 & 0.0407 &0.0765\\
		&our method & 30  & 0.0577 & 0.0412 & 0.0779 & 0.0579 & 0.0417 & 0.0791 & 0.0586 & 0.0422 & 0.0799\\
		\midrule
                 \multirow{3}{*}{ada} &no text &3 & 0.0594 & 0.0428 & 0.0799 & 0.0599 & 0.0436 & 0.0813 & 0.0602 &0.0436 & 0.0812\\
		& textblob &7    & 0.0565 & 0.0397 & 0.0751 & \bf 0.0565 &0.0397 & 0.0752 & 0.0566 & 0.0397 & \bf 0.0751\\
		&our method & 12 & \bf0.0564 & 0.0396 & \bf 0.0750 & \bf 0.0565 & 0.0396 & \bf 0.0751 & \bf 0.0565 & 0.0396 & \bf0.0751\\
		\bottomrule
	\end{tabular}
	      }
\end{table}

The results in Table~\ref{Comparison of Models for Crude Oil Price Forecasting} illustrate the forecasting performance of our proposed method and other benchmarks at horizons 1, 2 and 3. In this table, models named \emph {no text} are without any text features; models named \emph{textblob} contain our proposed shore text topic features and use the \emph{TextBlob} calculation results directly as the sentiment intensity; models named \emph{our method} contain our proposed short text topic and sentiment features. 
 The general \emph{arima} and \emph{arimax} model contains three parameters: \emph{p},\emph{d} and \emph{q}. \emph{p} is the number of autoregressive terms, \emph{q} is the number of sliding average terms, and \emph{d} is the number of differences taken to make the series stationary. We applied \emph{arima (p,d,q)} on \emph{no text} and 
The \emph{arimax (p,d,q)} on \emph{textblob} and \emph{our method}. The parameters are \emph{arima (4,0,3)}, \emph{arimax (4,1,3)}, and \emph{arimax (4,1,3)} for \emph{no text}, \emph{textblob}, and \emph{our method} respectively. The text features selected by \emph{our method} are listed in the Appendix \ref{appendix}, from which we can learn the preferences of different models for features. From the results, we can draw conclusions:

\begin{itemize}
\tightlist
\item[(1)] \textbf{Text vs no text}.  For \emph{rf}, \emph{svr}, and \emph{ada}, \emph{our method} performs better, which also proves that the attempt to add text features to the forecasting model is meaningful. On the other hand, text features are unlikely to improve forecasting accuracy  for \emph{arima}.
\item[(2)]  \textbf{Our sentiment index vs \emph{TextBlob}}. By comparing the results of \emph{textblob} and \emph{our method}, we find that \emph{our method} is slightly superior to the \emph{textblob} model in terms of \emph{svr} and \emph{ada}. This shows that our proposed sentiment index is better in certain circumstances.
\item[(3)] \textbf{Our method vs \emph{svr-Li} \citep{li2018text}}. \emph{svr-Li} is a forecasting model where the authors use multi-source text and financial features to forecast price. During this period our proposed method deeply extracts short textual features with less human intervention compared to \emph{svr-Li}. The forecasting improvement of our proposed approach is indeed caused by the several modifications including text source expansion, a more fine word embedding method \emph{Glove}, \emph{SeaNMF} short text topic model, continuous sentiment intensity, and the \emph{Adaboost.RT} model.
\item[(4)] \textbf{Optimal Model}.  The performance of \emph{ada} is generally better than the other models in terms of \emph{rmse} and \emph{mape}. It should be emphasized that the  \emph{arima} model without text features also performs well. We recommend \emph{arima} to users who can't get text features. For those who pursue higher forecasting accuracy, we suggest using the proposed method.
\end{itemize}

\subsection{DM significance tests}
In this part, we carry out Diebold-Mariano (DM) test \citep{harvey1997testing} to explore that if regression models combined with our proposed text indexes are significantly better or worse than methods like commonly used \emph{TextBlob} and \emph{svr-Li} \citep{li2018text}. The null hypothesis is that the two methods have the same forecast accuracy. The alternative hypothesis is that our method is less or more accurate than the standard method. Given a significance level $\alpha$ (eg,. 5\%), if the DM  test statistic falls in the lower or upper 2.5\% tail of a standard normal distribution, we reject the null hypothesis. The DM test is implemented using \texttt{forecast::dm.test()} in \textbf{R}. 
 
 \begin{table}[htbp]
	\normalsize
	\centering
	\caption{The entries show the p-values of DM tests that regression models combined with our proposed text indexes are better or worse than methods with commonly used \emph{textblob} and state-of-the-art \emph{svr-Li}  \citep{li2018text} over h=1, 2 and 3. If p-value \textless  2.5\%, we reject the null hypothesis and the number is bolded.}

	\label{median-seasonality-trend-accuracy}
	\resizebox{340pt}{35mm}{
		\begin{tabular}{lllllllll}
			\toprule
&&&\multicolumn{3}{c} {svr-Li}&\multicolumn{3}{c} {$textblob$}\\
\cmidrule(lr){4-6} \cmidrule(lr){7-9} 
			&&&h=1 &h=2&h=3&h=1&h=2&h=3   \\
			\cmidrule(lr){4-6} \cmidrule(lr){7-9} 		
\multirow{2}{*}{our text index} 	&rf&better   &1.0000&{1.0000}&{1.0000}&{0.5499}&{0.9219}&{0.9543}\\
	&	rf &worse&\textbf{0.0000}&\textbf{0.0000}&\textbf{0.0000}&0.4501 &0.0781&0.0457 \\
			\cmidrule(lr){4-6} \cmidrule(lr){7-9} 
	
\multirow{2}{*}{our text index} 	&svr&better   &\textbf{0.0121}&{0.1471}&\textbf{0.0203}&{0.2999}&{0.3466}&{0.4375}   \\

	&	svr &worse  &0.9879&0.8529&0.9797&0.7001 &0.6534&0.5625  \\
				\cmidrule(lr){4-6} \cmidrule(lr){7-9} 
	
\multirow{2}{*}{our text index} 	&arimax&better   &\textbf{0.0014}  &\textbf{0.0006}&0.9994&{0.8769}&{0.9389}&{1.0000}  \\

	&	arimax &worse  &0.9986&0.9993&\textbf{0.0006} &0.1231 &{0.0611}&\textbf{0.0000} \\
				\cmidrule(lr){4-6} \cmidrule(lr){7-9} 
	
\multirow{2}{*}{our text index} 	&ada&better   &\textbf{0.0000} &\textbf{0.0001}&\textbf{0.0000} &{0.7348}&{0.9986}&{0.8603}\\

	&	ada &worse  &1.0000&0.9999&1.0000&0.2652&\textbf{0.0014}&{0.1397}    \\
			
			\midrule
		\end{tabular}
		}
\end{table}

We can observe that 
\begin{itemize}
\tightlist
\item[(1)] Regression methods such as \emph{svr}, \emph{arimax} and \emph{ada} combined with our proposed index are significantly better than \emph{svr-Li} \citep{li2018text}, indicating our methods only use text features without considering any extra financial features, yielding better forecasts. 
\item[(2)] Results also show that our method is neither significantly better nor worse than \emph{textblob}. Our proposed topic intensity still needs to be re-designed and optimized in the future study.
\end{itemize}

Next, in order to identify the optimal model, we further want to validate if \emph{ada} with our proposed text index is significantly better than other regression methods. 
We conclude that 
\begin{itemize}
\tightlist
 \item[(1)] \emph{ada} combined with our proposed text index is significantly better than other regression models except for \emph{arimax} over h=1.
 \item[(2)]  On the other hand, it is interesting that \emph{ada} is not significantly worse than other all models.
 \end{itemize}

\begin{center}
\begin{table}[htbp]
	\normalsize
	\centering
	\caption{The entries show the p-values of DM tests that \emph{ada} combined with our proposed text index are better or worse than regression methods over h=1, 2 and 3. If p-value \textless 2.5\%, we reject the null hypothesis and the number is bolded. }
	\label{dm-test-for-ada-oil}
	\resizebox{340pt}{14mm}{
		\begin{tabular}{lllllllllll}
			\toprule
&&\multicolumn{3}{c} {rf}&\multicolumn{3}{c} {svr}&\multicolumn{3}{c} {arimax}\\
\cmidrule(lr){3-5} \cmidrule(lr){6-8} \cmidrule(lr){9-11} 
	&&h=1 &h=2&h=3&h=1&h=2&h=3&h=1&h=2&h=3 \\
			\cmidrule(lr){3-5} \cmidrule(lr){6-8}\cmidrule(lr){9-11} 
\multirow{2}{*}{ada}  &better &\textbf{0.0000}&\textbf{0.0000}&\textbf{0.0000}&\textbf{0.0001} &\textbf{0.0000} &\textbf{0.0001}&{0.0112}&\textbf{0.0051} &\textbf{0.0000}  \\
	 &worse&1.0000&1.0000&1.0000&0.9999&1.0000&0.9999&0.9888 &0.9949&{1.0000}\\
			
			\midrule
		\end{tabular}
		}
\end{table}
\end{center}

\section{Application to natural gas and gold price data}
\label{Application to Other Commodities Price Data}
In section~\ref{Text-driven crude oil price forecasting}, we briefly discussed the relationships among the three futures prices of crude oil, natural gas, and gold based on previous research. Since \emph{our method} incorporating news headlines can forecast crude oil prices pretty well, it's intuitive that it can also be migrated to other application scenarios. That is, \emph{our method} may be used to forecast the prices of natural gas and gold. It 
should be emphasized that since \citep{li2018text} only forecast crude oil prices and the experimental results in the previous section have verified that our method is more advanced than \emph{svr-li}, we do not include \emph{svr-li} in the comparison in this section for the price forecasts of natural gas and gold. And \emph{textblob} is no longer considered in this scenario either.

\subsection{Application to natural gas price data}
\label{Application to natural gas price data}

\begin{table}[htbp]
\footnotesize   
  \centering
  \caption{Optimal lag of 6 time series related to natural gas}
  \label{lag of gas}
  \resizebox{340pt}{12mm}{
    \begin{tabular}{llllllll}
      \toprule
      Time Series  & topic 1 &  topic 2 & topic 3 &topic 4 &polarity &dprice \\
      \midrule
      SIC & -12.4214 &-12.8220 & -12.3614 &-12.8161 & -13.6855 & -7.9601 \\
      lag  & 7 & 7 & 7 & 8 & 7 &3\\
      \bottomrule
    \end{tabular}
  }
\end{table}

\begin{table}[htbp]  
\footnotesize
	\centering
	\caption{Forecasting results of multiple methods for gas over h=1, 2 and 3.}  
	\label{Comparison of Models for Natural Gas Price Forecasting}
      \resizebox{460pt}{32mm}{
	\begin{tabular}{ccc ccc ccc ccc}
		\toprule
		&\multirow{2}{*}{Model} & Number of& \multicolumn{3}{c}{h=1} & \multicolumn{3}{c}{h=2} &\multicolumn{3}{c}{h=3} \\
		\cmidrule(lr){4-6} \cmidrule(lr){7-9} \cmidrule(lr){10-12} 
		 & & features & \emph{rmse} & \emph{mae} & \emph{mape} & \emph{rmse} &\emph{mae} & \emph{mape}  & \emph{rmse} &\emph{mae} & \emph{mape} \\
		\midrule
		\multirow{2}{*}{rf}& no text &3  & 0.0661 & 0.0430 & 0.1349 & 0.0662 & 0.0433 & 0.1382 & 0.0664 & 0.0428 & 0.1406 \\
               & our method &28       & 0.0630 & 0.0401 & 0.1355 &0.0582 & 0.0350 & 0.1237 & 0.0585 & 0.0349 & 0.1235   \\
		\midrule
              \multirow{2}{*}{svr}  & no text &3 & 0.0587 & 0.0354 & 0.1226 & 0.0592 & 0.0357 & 0.1237 & 0.0586 & 0.0352 & \bf 0.1223\\
		&our method &8       & 0.0585 & 0.0352 & 0.1237 &0.0582 & 0.0350& 0.1237 & 0.0585 & 0.0349 & 0.1235\\
		\midrule
                \multirow{2}{*}{arima \& arimax} & no text & --   & \bf 0.0581 & 0.0348 & 0.1222 & \bf 0.0581 & 0.0348 & \bf 0.1224 & \bf 0.0581 & 0.0348 & \bf 0.1223\\
		& our method & 30   & \bf 0.0581 & 0.0350& 0.1232 & 0.0585 &0.0357 & 0.1240 & 0.0587 & 0.0353 & 0.1231\\
		\midrule
                \multirow{2}{*}{ada} &no text & 3& \bf 0.0581 & \bf 0.0347 & 01224 & \bf 0.0581 & 0.0348 & 0.1225 &\bf 0.0581 & 0.0347 & 0.1225\\
		& our method & 19      &\bf  0.0581 & \bf 0.0347 & \bf 0.1222 & \bf 0.0581 & \bf 0.0347 & \bf 0.1224 & \bf 0.0581 & \bf 0.0346 & 0.1224 \\
		\bottomrule
	\end{tabular}
	      }
\end{table}

Analogously, we first calculate the lag of the natural gas-related time series, as shown in Table~\ref{lag of gas}. The parameters of \emph{p, d, q} are \emph{arima (3,0,4)} for \emph{no text} and \emph{arimax (4,1,2)} for \emph{our method}. From Table~\ref{Comparison of Models for Natural Gas Price Forecasting}, conclusions similar to those in Section~\ref{Feature selection and forecasting} can be obtained. For the \emph{rf}, \emph{svr} and \emph{ada} , our approach shows its superiority over \emph{no text} and performs best on \emph{ada}. However, the \emph{arimax} model, which uses our proposed text features, does not significantly outperform the \emph{arima} model.

\subsection{Application to gold price data}
\label{Application to gold price data}

\begin{table}[htbp]
\footnotesize   
  \centering
  \caption{Lag of 6 time series related to gold}
  \label{lag of gold}
  \resizebox{340pt}{12mm}{
    \begin{tabular}{llllllll}
      \toprule
      Time Series  & topic 1 &  topic 2 & topic 3 &topic 4 &polarity &dprice \\
      \midrule
      SIC & -12.6794 & -12.6277 & -12.6198  & -13.0770 & -13.9493 & -8.2163 \\
      lag  & 7 & 7 & 7 & 7 & 7 &4\\
      \bottomrule
    \end{tabular}
  }
\end{table}

\begin{table}[htbp]  
\footnotesize
	\centering
	\caption{Forecasting results of multiple methods based on these key factors for gold over h=1, 2 and 3.}  
	\label{Comparison of Models for gold Price Forecasting}
      \resizebox{460pt}{32mm}{
	\begin{tabular}{cc cccc cccc cccc}
		\toprule
		&\multirow{2}{*}{Model} & Number of& \multicolumn{3}{c}{h=1} &  \multicolumn{3}{c}{h=2} &\multicolumn{3}{c}{h=3} \\
		\cmidrule(lr){4-6} \cmidrule(lr){7-9} \cmidrule(lr){10-12} 
		 & & features & \emph{rmse} & \emph{mae} & \emph{mape} & \emph{rmse} &\emph{mae} & \emph{mape}  & \emph{rmse} &\emph{mae} & \emph{mape} \\
		\midrule
		\multirow{2}{*}{rf} & no text & 4  & 0.0528& 0.0367 & 0.0562 &0.0539 & 0.0385 & 0.0591 &0.0516 & 0.0367 & 0.0563  \\
               & our method & 40  & 0.0507 & 0.0353 & 0.0545 & 0.0517 & 0.0362 & 0.0558 & 0.0504 & 0.0359 & 0.0555  \\
		\midrule
                \multirow{2}{*}{svr} & no text &4 & 0.0452 & 0.0299 & 0.0460 & 0.0460 & 0.0308 & 0.0474 & 0.0456 & 0.0302 & 0.0465\\
		& our method &4      & 0.0467 & 0.0319 & 0.0498 & 0.0471 & 0.0323 & 0.0504 & 0.0474 & 0.0323 & 0.0504\\
		\midrule
                \multirow{2}{*}{arima} & no text &--    &0.0449 & \bf 0.0292 & 0.0451 & \bf 0.0449 & \bf 0.0293 & 0.0452 & 0.0449 &\bf 0.0293& \bf 0.0451 \\
		&our method &11 &0.0450 &0.0296 &0.0456 &0.0454 & 0.0303 & 0.0463 & 0.0452 &0.0296 &0.0459\\
		\midrule
               \multirow{2}{*}{ada}& no text & 4 & 0.0449 & \bf 0.0292 & \bf 0.0450 & \bf 0.0449 & \bf 0.0293 & \bf 0.0451 &0.0449 & \bf 0.0293 &\bf 0.0451\\
		&our method &15 &  \bf 0.0447 & \bf 0.0292 & 0.0451 & \bf 0.0449 & \bf 0.0293 & \bf 0.0451 & \bf 0.0447 & \bf 0.0293 & \bf 0.0451  \\
		\bottomrule
	\end{tabular}
	      }
\end{table}



The parameters of \emph{p,d,q} are \emph{arima (4,1,3)} for \emph{no text} and \emph{arimax (3,2,1)} for \emph{our method}.  From Table~\ref{Comparison of Models for gold Price Forecasting}, our method only outperforms \emph{no text} on \emph{rf}. Thus, it can be seen that the proposed textual features have a less significant role in improving the forecasting accuracy of gold prices. We conjecture that this is due to the low discussion of gold in all news text.

\section{Discussion}
\label{Discussion}
The international crude oil prices are influenced by many external factors in addition to historical price fluctuations, making the accurate and reliable forecasting a difficult task. The empirical results in \citet{Demirer2010The} and \citet{Kaiser2010The} suggest that incorporating UCG information from the social media into the crude oil price forecasting can achieve better performance. In order to obtain higher accuracy, researchers are free to add more various exogenous variables to their forecasting framework, such as \citet{li2018text}.
Although these exogenous variables can boost the forecasting ability to some extent, they also bring some uncertainty in forecasting due to excessive human intervention. What's more, their work does not systematically examine whether textual features or these exogenous features lead to good predictions. In this work, we suggest a change from adding more exogenous variables to construct high quality features from short and sparse text and mitigate the importance of manual intervention.
Our study focuses on extracting as much information as possible from news headlines to assist in forecasting crude oil prices, without manually choosing many other exogenous variables, yielding comparable performance.

Specifically, to fully extract features from short and sparse news headlines, we employ advanced \emph{GloVe} instead of \emph{bag of words} during word embedding as \emph{GloVe}'s pretrained model makes full use of massive corpus information, retains more semantic relationships, and saves considerable time while  \emph{bag of words} focuses more on syntax than on semantics. 

For topic modelling, \emph{LDA} is designed for accommodating long text and has been used in \citet{li2018text}. However, news headlines are short and sparse, making traditional \emph{LDA} difficult to discover potential topics from them due to the lack of contextual information \citep{shi2018short}. This mismatch between the text and topic model directly has a bad influence on the forecasting performance. Considering that, we employ advanced short text model \emph{SeaNMF} to construct topic intensity, with the aim at tackling the short and sparse news headlines and thus improving the forecasting performance. Our empirical results also show that \emph{SeaNMF} is more suitable to infer potential topics for the short and sparse news headlines than \emph{LDA}.

In terms of sentiment analysis, the sentiment indicator constructed in \citet{li2018text} is static and simple, ignoring the dynamic relationship with the previous days. So we design a novel sentiment indicator, taking the cumulative and diminishing effect of the market into consideration. 

Our empirical results prove that the proposed two marketing indicators are systematically combined with other key factors and thus produce more accurate forecasts compared with \citep{li2018text}. DM significance tests show that regression models such as \emph{svr}, \emph{arimax} and \emph{ada} combined with our proposed text indexes are significantly better than state-of-the-art \emph{svr-Li} \citep{li2018text}, indicating that our methods produce better forecasts with fewer human intervention.
Also, it is interesting that our proposed sentiment indicator is neither significantly better or worse than methods combined with \emph{TextBlob}. 
We further verify the superiority of \emph{ada} combined with the proposed text indicators with others.

Another significant merit is that our forecasting framework can also yield good forecasting performance when applied to other futures commodities, reflecting the flexibility and robustness of it. Due to the use of futures-related news headlines as an experimental training corpus, our method also obtains the expected good results in forecasting the prices of natural gas and gold. In future, the research framework can be transferred to other fields. For example, the news text features of listed companies can be added to the model to enhance the accuracy of its stock price prediction.

The limitation of this study is that it extracts text features from two dimensions, topic and sentiment, which can be further fully obtained by adding additional perspectives such as news categories, distribution and density. In the future, comments on the news from investors or other people who pay attention to crude oil could also be considered in our framework.


\section{Concluding remarks}
\label{Conclusion and future work}
Inspired by the work \citep{li2018text}, where they construct text indicators from news headlines and add some exogenous financial features to their forecasting model, we have reproduced their experimental process, studied their ideas in depth, and proposed some modifications and innovations. To improve forecasting performance, we particularly focus on the modelling for sparse and short news headlines. Two novel indicators based on sparse and short text are combined with other models and produce good performance. Applying the proposed approaches in natural gas and gold price forecasting is a strong support of the validity and generalizability of the research.

Our research framework provides an automated tool for crude oil price forecasting. As far as its practical application is concerned, our method is more suitable for price forecasting with a large amount of historical data. It is even better if there are corresponding news in the past period, as text features can be extracted from them to help improve the prediction. Also, our research focuses more on extracting text features from two dimensions, topic and sentiment, which can be further fully obtained by adding additional perspectives such as news categories, distribution and density.

\section*{Acknowledgements}

We are grateful to the Editor, two anonymous reviewers for
their helpful comments that improved the contents of this paper. We are also grateful to Professor Yanfei Kang from Beihang Univeristy for her meaningful suggestions and insights of this paper.

\section* {Appendix}
\label{appendix}
{Experimental setup and feature selection results for the crude oil forecasting} 
\begin{itemize}
\tightlist

\item Parameters for random forest regression: $min\_samples\_split=2$, $min\_samples\_leaf=1$, $min\_weight\_fraction\_leaf=0.0$, $max\_features=auto $;

\item Parameters for support vector regression: $kernel=sigmoid$, $max\_iter=100$;

\item Parameters for AdaBoost.RT: $n\_estimators=30$, $learning\_rate=0.01$, and the base estimator is ``DecisionTreeRegressor".
\end{itemize}

\label{Feature selection results: one step}
\begin{table}[htbp]
  \centering
  \caption{Feature selection results for crude oil price forecasting}
  \resizebox{320pt}{120mm}{
    \begin{tabular}{lcccc}
    \toprule
    features & \multicolumn{1}{l}{\emph{rf-text} (27)} & \multicolumn{1}{l}{\emph{svr-text} (17)} & \multicolumn{1}{l}{\emph{arimax} (4,1,3) (3)} & \multicolumn{1}{l} {\emph{ada-text} (12)} \\
    \midrule
    topic1(t-7) & \checkmark     & \checkmark      &       &  \\
    topic1(t-6) &       &       &      &  \\
    topic1(t-5) &       &       &    &  \\
    topic1(t-4) &   \checkmark       &       &       &  \\
    topic1(t-3) & \checkmark     & \checkmark     &      & \checkmark \\
    topic1(t-2) &       &       &    &  \\
    topic1(t-1) & \checkmark     & \checkmark     &     &  \checkmark  \\
    topic2(t-7) & \checkmark     &       &      &    \\
    topic2(t-6) &       &       &       &   \\
    topic2(t-5) & \checkmark     &       &       &  \\
    topic2(t-4) & \checkmark     & \checkmark     &     &  \checkmark  \\
    topic2(t-3) & \checkmark     &       &      &  \\
    topic2(t-2) &      & \checkmark     &       &  \\
    topic2(t-1) & \checkmark     & \checkmark     &  \checkmark      & \checkmark \\
    topic3(t-7) & \checkmark     &       &     &  \checkmark  \\
    topic3(t-6) & \checkmark     &  \checkmark     &      & \\
    topic3(t-5) &      &       &      &  \\
    topic3(t-4) & \checkmark     &    &       &  \\
    topic3(t-3) & \checkmark     &      &       &  \\
    topic3(t-2) & \checkmark     &\checkmark       &       &  \\
    topic3(t-1) & \checkmark     &       &   &  \\
    topic4(t-7) &       &       &       &  \\
    topic4(t-6) &       &       &       &  \\
    topic4(t-5) & \checkmark     & \checkmark     &     & \checkmark \\
    topic4(t-4) & \checkmark     &       &       &  \\
    topic4(t-3) & \checkmark     & \checkmark     &       &  \\
    topic4(t-2) &  \checkmark     &       &       &  \\
    topic4(t-1) &      &       &       &  \\
    polarity(t-7) & \checkmark     & \checkmark      &       & \checkmark \\
    polarity(t-6) & \checkmark     & \checkmark     &       &  \checkmark  \\
    polarity(t-5) & \checkmark     & \checkmark     &  \checkmark       & \checkmark \\
    polarity(t-4) &       &       &       &  \\
    polarity(t-3) & \checkmark     & \checkmark     &       &  \\
    polarity(t-2) &       &       &       &  \\
    polarity(t-1) & \checkmark     &      &       &  \\
    dprice(t-3) & \checkmark     & \checkmark     &       &  \checkmark  \\
    dprice(t-2) & \checkmark     & \checkmark     &       & \checkmark \\
    dprice(t-1) & \checkmark     & \checkmark     & \checkmark      & \checkmark \\
    \bottomrule
    \end{tabular}%
   }
  \label{tab:addlabel}%
\end{table}%

\begin{table}[htbp]
  \centering
  \caption{Feature selection results for natural gas forecasting}
  \resizebox{320pt}{120mm}{
    \begin{tabular}{lcccc}
    \toprule
   features & \multicolumn{1}{l}{\emph{rf-text} (28)} & \multicolumn{1}{l}{\emph{svr-text} (8)} & \multicolumn{1}{l}{\emph{arimax} (4,1,2)(30)} & \multicolumn{1}{l}{\emph{ada-text} (19)} \\
    \midrule
    topic1(t-7) &\checkmark       &       & \checkmark     &  \\
    topic1(t-6) &       &       &     &  \\
    topic1(t-5) & \checkmark     &\checkmark       & \checkmark     &\checkmark  \\
    topic1(t-4) &       &       &     &  \\
    topic1(t-3) & \checkmark      &       & \checkmark     &  \\
    topic1(t-2) & \checkmark      &       & \checkmark      &  \\
    topic1(t-1) &       &       &      &  \\
    topic2(t-7) & \checkmark     &       & \checkmark     & \checkmark \\
    topic2(t-6) & \checkmark     &       & \checkmark     &  \\
    topic2(t-5) &\checkmark       &       & \checkmark     &  \\
    topic2(t-4) &  \checkmark     &       & \checkmark     &  \\
    topic2(t-3) & \checkmark     &       & \checkmark     & \checkmark \\
    topic2(t-2) & \checkmark     &       & \checkmark     &\checkmark  \\
    topic2(t-1) &       &       & \checkmark     &  \\
    topic3(t-7) &  \checkmark     & \checkmark      & \checkmark     & \checkmark \\
    topic3(t-6) & \checkmark     &    & \checkmark     & \checkmark \\
    topic3(t-5) &       &       &    &  \\
    topic3(t-4) &       &       &       &  \\
    topic3(t-3) & \checkmark      &       & \checkmark     & \checkmark \\
    topic3(t-2) &       &       &     &  \\
    topic3(t-1) & \checkmark     &  \checkmark     & \checkmark     & \checkmark \\
    topic4(t-8) & \checkmark     & \checkmark     & \checkmark     & \checkmark \\
    topic4(t-7) &\checkmark       &       & \checkmark     &  \\
    topic4(t-6) &  \checkmark     &       & \checkmark     & \checkmark \\
    topic4(t-5) &       &       & \checkmark     & \checkmark \\
    topic4(t-4) & \checkmark     &       & \checkmark     &  \\
    topic4(t-3) &  \checkmark     &       & \checkmark     &  \\
    topic4(t-2) &       &       &      &  \\
    topic4(t-1) & \checkmark     &       & \checkmark     &  \\
    polarity(t-7) & \checkmark     & \checkmark     & \checkmark     & \checkmark \\
    polarity(t-6) & \checkmark     &       & \checkmark     & \checkmark \\
    polarity(t-5) & \checkmark     &       & \checkmark     & \checkmark \\
    polarity(t-4) &       &       &    &  \\
    polarity(t-3) & \checkmark     &       & \checkmark     & \checkmark \\
    polarity(t-2) &       &       &       &  \\
    polarity(t-1) & \checkmark     & \checkmark      & \checkmark     & \checkmark \\
    dprice(t-3) & \checkmark     & \checkmark     & \checkmark     & \checkmark \\
    dprice(t-2) & \checkmark     &       & \checkmark     & \checkmark \\
    dprice(t-1) & \checkmark     & \checkmark     & \checkmark     & \checkmark \\
    \bottomrule
    \end{tabular}%
  }
  \label{tab:addlabel}%
\end{table}%

\begin{table}[htbp]
  \centering
  \caption{Feature selection results for gold price forecasting}
 \resizebox{320pt}{120mm}{
    \begin{tabular}{lcccc}
    \toprule
    features & \multicolumn{1}{l}{\emph{rf-text} (31)} & \multicolumn{1}{l}{\emph{svr-text} (3)} & \multicolumn{1}{l}{\emph{arimax} (3,2,1) (24)} & \multicolumn{1}{l}{\emph{ada-text} (5)} \\
    \midrule
    topic1(t-7) & \checkmark     &       & \checkmark     &  \\
    topic1(t-6) & \checkmark     &       & \checkmark     &  \\
    topic1(t-5) & \checkmark     &       & \checkmark     &  \\
    topic1(t-4) & \checkmark     &       & \checkmark     &  \\
    topic1(t-3) & \checkmark     &       &       &  \\
    topic1(t-2) &       &       &       &  \\
    topic1(t-1) & \checkmark     &       &       &  \\
    topic2(t-7) & \checkmark     &       & \checkmark     &  \\
    topic2(t-6) & \checkmark     &       &       &  \\
    topic2(t-5) & \checkmark     &       & \checkmark     &  \\
    topic2(t-4) &       &       &       &  \\
    topic2(t-3) & \checkmark     &       & \checkmark     & \checkmark \\
    topic2(t-2) & \checkmark     &       &       &  \\
    topic2(t-1) & \checkmark     &       & \checkmark     &  \\
    topic3(t-7) & \checkmark     &       & \checkmark     &  \\
    topic3(t-6) &       &       & \checkmark     &  \\
    topic3(t-5) &       &       &       &  \\
    topic3(t-4) &       &       & \checkmark     &  \\
    topic3(t-3) & \checkmark     &       &       &  \\
    topic3(t-2) & \checkmark     &       &       &  \\
    topic3(t-1) & \checkmark     &       &       &  \\
    topic4(t-7) & \checkmark     &       & \checkmark     &  \\
    topic4(t-6) & \checkmark     &       & \checkmark     &  \\
    topic4(t-5) &       &       &       &  \\
    topic4(t-4) & \checkmark     &       &       & \checkmark \\
    topic4(t-3) & \checkmark     &       & \checkmark     &  \\
    topic4(t-2) & \checkmark     &       &       &  \\
    topic4(t-1) & \checkmark     & \checkmark     & \checkmark     & \checkmark \\
    polarity(t-7) & \checkmark     &       & \checkmark     &  \\
    polarity(t-6) & \checkmark     &       & \checkmark     &  \\
    polarity(t-5) & \checkmark     &       & \checkmark     &  \\
    polarity(t-4) & \checkmark     &       &       &  \\
    polarity(t-3) &       &       & \checkmark     &  \\
    polarity(t-2) & \checkmark     &       & \checkmark     &  \\
    polarity(t-1) &       &       &       &  \\
    dprice(t-4) & \checkmark     &       & \checkmark     &  \\
    dprice(t-3) & \checkmark     &       & \checkmark     &  \\
    dprice(t-2) & \checkmark     & \checkmark     & \checkmark     & \checkmark \\
    dprice(t-1) & \checkmark     & \checkmark     & \checkmark     & \checkmark \\
    \bottomrule
    \end{tabular}%
  }
  \label{tab:addlabel}%
\end{table}%

\biboptions{longnamesfirst}
\bibliography{crude-oil-forecasting-revise}
\end{document}